\documentclass[journal]{IEEEtran}
\usepackage{cite}
\usepackage{amsmath}
\usepackage{multirow}
\usepackage{graphicx}
\usepackage{booktabs}

\ifCLASSINFOpdf
\else
   \usepackage[dvips]{graphicx}
\fi
\usepackage{url}

\usepackage{graphicx}

\begin{document}

\title{XLST: Cross-lingual Self-training to Learn Multilingual Representation for Low Resource Speech Recognition}

\author{Zi-Qiang Zhang, Yan Song, Ming-Hui Wu, Xin Fang, Li-Rong Dai
\thanks{Z.-Q. Zhang, Y. Song, X. Fang and L.-R. Dai are with the National Engineering Laboratory
	for Speech and Language Information Processing, University of Science and
	Technology of China, Hefei 230027, China, (e-mail: zz12375@mail.ustc.edu.cn; songy@ustc.edu.cn; klg@mail.ustc.edu.cn; lrdai@ustc.edu.cn).}
}
\maketitle

\begin{abstract}
In this paper, we propose a weakly supervised multilingual representation learning framework, called cross-lingual self-training (XLST).
XLST is able to utilize a small amount of annotated data from high-resource languages to improve the representation learning on multilingual un-annotated data.
Specifically, XLST uses a supervised trained model to produce initial representations and another model to learn from them, by maximizing the similarity between output embeddings of these two models.
Furthermore, the moving average mechanism and multi-view data augmentation are employed, which are experimentally shown to be crucial to XLST. 
Comprehensive experiments have been conducted on the CommonVoice corpus to evaluate the effectiveness of XLST.
Results on 5 downstream low-resource ASR tasks shows that our multilingual pretrained model achieves relatively 18.6\% PER reduction over the state-of-the-art self-supervised method, with leveraging additional 100 hours of annotated English data.
\end{abstract}

\begin{IEEEkeywords}
Multilingual representation learning, cross-lingual self-training, low-resource speech recognition.
\end{IEEEkeywords}

\IEEEpeerreviewmaketitle

\section{Introduction}
Modern automatic speech recognition (ASR) systems are usually trained on hundreds of hours of annotated data~\cite{li2020comparison,wang2020transformer}, which is not available for most of low-resource languages~\cite{besacier2014automatic}.
To tackle the low-resource ASR problem, pretraining techniques have shown to be promising, where effective speech representations are first learned from other available materials (e.g. out-of-language data or un-annotated speech data), then transferring to the target low-resource language.

Existing pretraining methods can be conducted in either supervised or unsupervised manners.
Early researches focus on supervised pretraining with annotated data of one or multiple high-resource languages~\cite{stolcke2006cross,thomas2010cross,vesely2012language,thomas2012multilingual,lal2013cross,thomas2013deep,huang2013cross,ghoshal2013multilingual,vu2014multilingual,sercu2016very,cho2018multilingual,dalmia2018sequence,tong2018cross,karafiat2019analysis}.
The pretrained model is used either as the feature extractor~\cite{stolcke2006cross,thomas2010cross,thomas2012multilingual,vesely2012language,lal2013cross,thomas2013deep}, or to initialize the target acoustic model in hybrid~\cite{huang2013cross,ghoshal2013multilingual,vu2014multilingual,sercu2016very} and end-to-end~\cite{cho2018multilingual,dalmia2018sequence,tong2018cross,karafiat2019analysis} systems.

Compared with supervised methods, unsupervised pretraining has the advantage of employing large amount of un-annotated speech data.
In this scenario, an unsupervised model is expected to learn meaningful representations from speech, ideally carrying phonetic structures like a supervised model.
For example, it could be trained by reconstructing input frames~\cite{renshaw2015comparison,chorowski2019unsupervised,Chung2019,liu2020mockingjay,liu2020tera,ling2020decoar} in a masked or auto-regressive manner.
Recently, self-supervised methods catch much interests in speech community~\cite{oord2018representation}.
Contrastive loss implemented by distinguishing between positive and negative samples is proposed there to learn more discriminative representations as well as preventing model collapse~\cite{Schneider2019,baevski2019vq,Baevski2020wav2vec2}.
Among them, Wav2vec 2.0~\cite{Baevski2020wav2vec2} is one of the state-of-the-art methods both on in-language and cross-lingual pretraining~\cite{conneau2020unsupervised}.
Despite its promising results, Wav2vec 2.0 framework is computationally expensive~\cite{Baevski2020wav2vec2}.

For a specific language, there is another way to utilize both annotated and un-annotated data called self-training~\cite{Kahn2020self,Park2020Improved} or teacher-student learning~\cite{li2014learning,Li2017Large,meng2019Domain}.
A teacher model is first trained on an annotated speech dataset of the target language to produce initial pseudo labels for a larger un-annotated speech dataset of the same language, and then a student model is trained with such pseudo-annotated data.
Though self-training is capable of making full use of both annotated and un-annotated data, it is not designed specifically for representation learning, and is a kind of training within one language where both annotated and un-annotated data comes from the same language.

In this work we consider improving multilingual representation learning by leveraging a certain amount of annotated data from high-resource languages.
Inspired by self-training, we propose the Cross-lingual Self-training (XLST) as a new kind of multilingual pretraining framework.
XLST supposes that frame-level acoustic representations could be shared in some degree across different languages.
Its training procedure is similar to the usual self-training with the main difference that the annotated and un-annotated data comes from different languages, \textit{i.e.} annotated data from high-resource languages and un-annotated data from low-resource languages.
Furthermore, instead of producing pseudo labels, XLST generates frame embeddings as training targets.
Inspired by~\cite{grill2020bootstrap,chen2020simple}, we implement XLST with parallel networks, one for producing targets and another for acoustic modeling.
They are tied in a moving average (MA) way~\cite{grill2020bootstrap} so that the produced targets could be refined online.
These two networks take different augmented views of a same input and the loss is computed as their output dissimilarity.

Our work is also related to a very recent method, Unispeech~\cite{wang2021unispeech}, which also utilized annotated data of high-resource languages to improve multilingual representations.
While they used all data simultaneously by multi-task learning under Wav2vec 2.0~\cite{Baevski2020wav2vec2} framework and here we employ the proposed XLST.

Our contributions include: 
1) we propose XLST, a multilingual representation learning framework capable to use both annotated data from high-resource languages and multilingual un-annotated data;
2) We show experimentally multi-view augmentation and online moving average (MA) are crucial to XLST;
3) XLST learns better multilingual representations than self-supervised methods, demonstrated by significant performance improvement on downstream speech recognition tasks.

\section{Methodology}
\subsection{General formulation of the representation learning problem} \label{ssec:formulation}
Given an unlabeled dataset $\mathcal{D}^u$ with $N$ samples $\left\{\boldsymbol{x}^1,\boldsymbol{x}^2,\ldots, \boldsymbol{x}^N\right\}$, representation learning aims to tune a model $f_\theta$ that can encode the input $\boldsymbol{x}$ into a high-level structured embedding $\boldsymbol{e}=f_\theta\left(\boldsymbol{x}\right)$.

In self-supervised representation learning, the model is trained by maximizing the output similarity from different augmented versions of the input sample, which can be formulated as 
\begin{equation}
\mathcal{L}=\boldsymbol{sim}\left(f_\theta\left(A\left(\boldsymbol{x}\right)\right), g_\phi\left(A'\left(\boldsymbol{x}\right)\right)\right) \label{eq:loss1}
\end{equation}
where $A$ and $A'$ indicate random data augmentations, and $g_\phi$, $f_\theta$ denote the models with parameters $\phi$ and $\theta$ respectively.
$g_\phi$ could be considered as a target producer for the main model $f_\theta$.
In visual tasks, $g_\phi$ could simply be $f_\theta$, \textit{i.e.} $g=f$ and $\phi=\theta$~\cite{chen2020simple, chen2020exploring}, or with the same network but tied parameters~\cite{grill2020bootstrap}.
In speech tasks, $g$ could be defined as a front part of $f$, specifically a stacked CNN in~\cite{Baevski2020wav2vec2}.

Self-supervised methods may face the ``model collapse'' issue, where $f_\theta$ and $g_\phi$ are independent of input~\cite{grill2020bootstrap,chen2020exploring}.
Their outputs are the same but obviously no meaningful representations could be learned.
Existing speech representation learning methods prevent this issue by using contrastive loss to replace Eqn.~\ref{eq:loss1}~\cite{Schneider2019,baevski2019vq,Baevski2020wav2vec2}.

\subsection{XLST for multilingual speech representation learning} \label{ssec:XLST}
\begin{figure}
	\centerline{\includegraphics[width=\columnwidth]{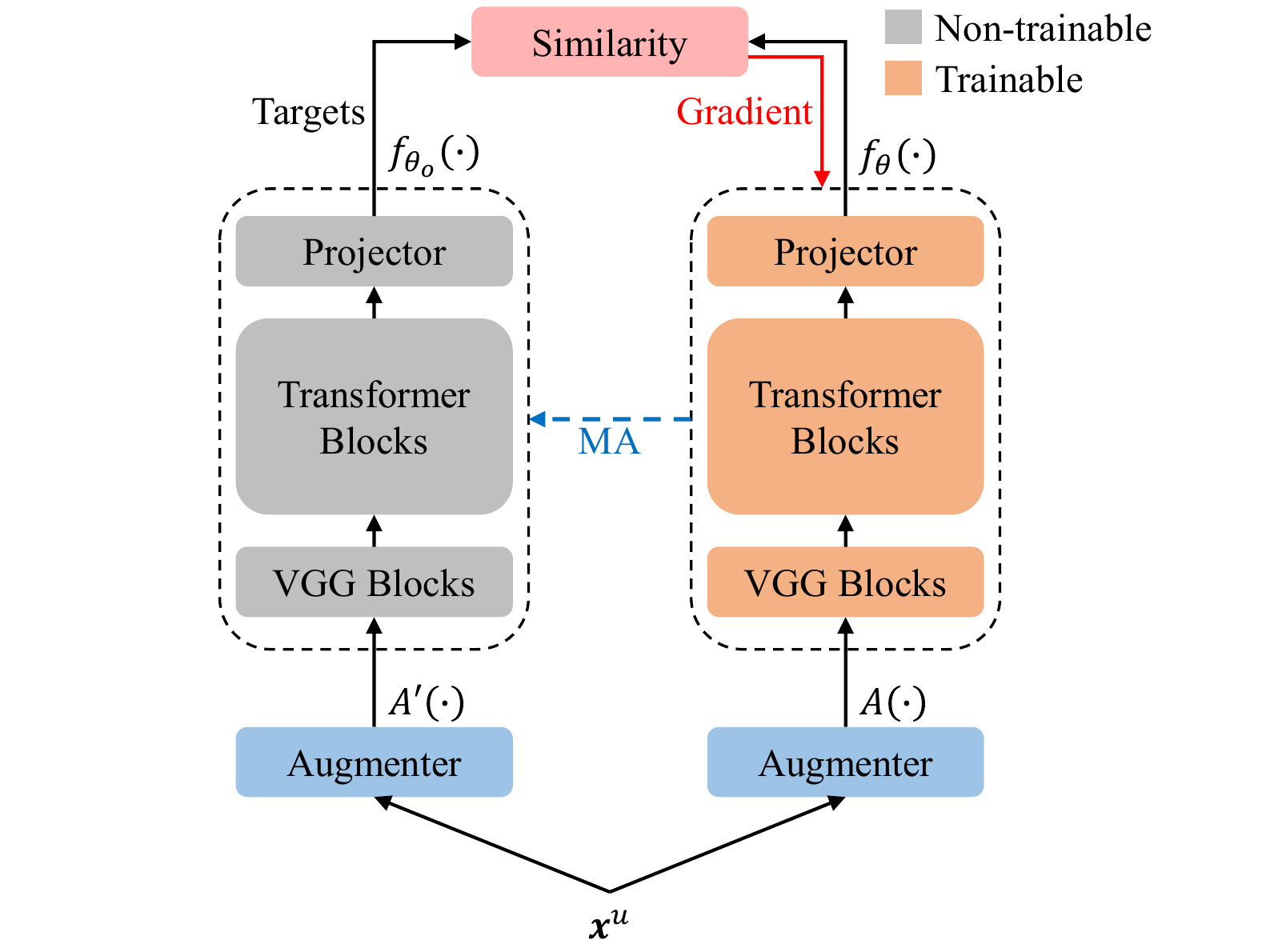}}
	\caption{XLST model architecture. The \textit{Target Network} ($f_{\theta_o}$) produces frame-level targets while the \textit{Main Network} ($f_\theta$) is trained to maximize the similarity between its output and the targets. Two parallel networks are tied in a moving average way. $A$ and $A'$ indicate different random augmentation on one multilingual sequence input $\boldsymbol{x}^u$. }
	\label{fig:XLST-1}
\end{figure}

This paper considers multilingual speech scenario, where there are $M$ un-annotated datasets $\left\{\mathcal{D}^1,\mathcal{D}^1,\ldots,\mathcal{D}^M\right\}$ from $M$ languages.
We inherit the basic idea of Eqn.~\ref{eq:loss1} with the difference that we adopt a external trained model (denoted as the \textit{Target Network} in Fig.~\ref{fig:XLST-1}) as the target producer, acting like a teacher model in self-training.
Initialization of the target producer will be detailed in the next subsection.
In our way, training could be simple without using contrastive loss and the learned representations could benefit from the prior knowledge provided by the external model.

The architecture of our proposed multilingual pretraining framework, XLST, is illustrated in Fig.~\ref{fig:XLST-1}.
XLST contains two parallel networks, a non-trainable \textit{Target Network} ($f_{\theta_o}$) and a trainable \textit{Main Network} ($f_\theta$).
The $f_{\theta_o}$ and $f_\theta$ are designed with the same architecture, composed of a stack of CNN Blocks, Transformer Blocks and a nonlinear Projector.
The input sequence $\boldsymbol{x}^u \in \mathcal{R}^{T\times F}$ ($T$ is the sequence length and $F$ is the dimension of acoustic features) is first preprocessed by a random augmenter $A(\cdot)$, to provide multiple views of the same sequence.
The augmenter can be implemented by random masking consecutive frames and frequency bins, or mixing the input with another sentence~\cite{zhang2017mixup,verma2019interpolation}.
It encourages the model to learn context dependency in both time and frequency dimensions.

Let $\boldsymbol{e}$ and $\boldsymbol{z}$ denote the output embeddings from the \textit{Target Network} and the \textit{Main Network} respectively, the similarity loss can be computed using normalized frame-level squared Euclidean distance similar to~\cite{grill2020bootstrap}, 
\begin{equation}
\boldsymbol{e} = f_{\theta}\left(A\left(\boldsymbol{x}^u\right)\right);\quad \boldsymbol{z} = f_{\theta_o}\left(A'\left(\boldsymbol{x}^u\right)\right) \label{eq:forward}
\end{equation}
\begin{equation}
\mathcal{L}_{sim}\left(\boldsymbol{e}, \boldsymbol{z}\right) = \sum_{i=1}^{T} ||\overline{z}_i - \overline{e}_i||^2
= \sum_{i=1}^{T} 2 - 2\frac{\left<z_i, e_i\right>}{||z_i||||e_i||} \label{eq:loss}
\end{equation}
where $\boldsymbol{z}=z_1,z_2,\ldots,z_T$ and $\boldsymbol{e}=e_1,e_2,\ldots,e_T$ are frame level embeddings with projected dimension of $d$.

\subsection{Target network initialization and refining} \label{ssec:initialization}
Though in visual tasks like~\cite{grill2020bootstrap} even random initialization of the \textit{Target Network} could result in good performance, our primary experiments indicate it is not a good choice for speech tasks.
In this work the \textit{Target Network} is first trained on a provided annotated $\mathcal{D}^o$ from a high-resource language, e.g. English.
An linear classifier is added on the top of the \textit{Target Network} to compute frame-level probabilities w.r.t. the phoneme inventory of this language.
During this stage, the augmenter mentioned above could also be applied.

Due to the limit of annotated data and language mismatch, it is expected that the initial targets produced by the \textit{Target Network} could not perform well on multilingual data.
Therefore, the \textit{Target Network} is refined during training, producing better quality of targets.
It could be done by iteratively assigning the parameters of the \textit{Main Network} to the \textit{Target Network}, or in a online way.
In this work we use the online moving average (MA)~\cite{grill2020bootstrap} with moving coefficient $\lambda$, it happens after every optimization step of the \textit{Main Network}
\begin{equation}
\theta_o \leftarrow \lambda \theta_o + \left(1-\lambda\right)\theta \label{eq:moving}
\end{equation}

\section{Experimental Setup}
In this section, we conduct preliminary experiments to evaluate the performance of XLST for multilingual ASR.
Specifically, we want to answer the following questions:
1) Starting with a model of one non-target high-resource language, can XLST learn good representations from multilingual un-annotated data?
2) What is important to the effectiveness of XLST?

\noindent{\textbf{Data and model configuration.}}
Since annotated data usually tends to be less than un-annotated data, we use 100 hours of English annotated speech and up to 793 hours of un-annotated speech from 10 languages.
The English data is from the Librispeech corpus~\cite{Panayotov2015Librispeech}\footnote{We use the aligned frame-level phoneme labels for Librispeech-100h provided by~\cite{oord2018representation}.}
and the un-annotated data is from the CommonVoice corpus~\cite{ardila2020common}\footnote{As the older release version is unavailable now, we use the December 2019 release version and keep the same hours of data as~\cite{conneau2020unsupervised} for 10 languages.}.
For comparison, 1250 hours of English un-annotated speech from the CommonVoice dataset is also used.
Note that we use the same amount of un-annotated data for every language as~\cite{conneau2020unsupervised}, with only the difference that we replace 100 hours of English data with annotated Librispeech-100h, which is less than 8\% of the total data (1350h un-annotated data used in~\cite{conneau2020unsupervised}).

Models are trained with different data configurations, shown in Table~\ref{Tab:general results}.
Specifically, 1) T-100, supervised baseline, which is also used as the initial target producer. 
2) ST-xx series, English pretrained models with different dataset sizes.
ST-en457 is trained on a 457-hour subset and ST-en1250 is trained on the whole English dataset.
The \textit{Target Network} initialization of these two models can be considered as an offline iterative optimization strategy.
3) XLST-mono, models pretrained on the un-annotated dataset of each language individually.
To show whether the language-specific representations can be learned from an English model, ST-en1250 is used as the initial \textit{Target Network}.
4) XLST-multi, the final multilingual pretrained model.
ST-en457 is used as the initial \textit{Target Network} so that the total data amount is comparable to~\cite{conneau2020unsupervised}.
We also use the same data balance technique as~\cite{conneau2020unsupervised}.

\noindent\textbf{Evaluation.}
The effectiveness of pretrained models are evaluated by fine-tuning on downstream phoneme recognition tasks.
Following~\cite{riviere2020unsupervised,conneau2020unsupervised}, a linear classifier concatenating 2 successive frames (40ms) is added on the top of the \textit{Main Network}.
And the CTC loss~\cite{graves2006connectionist} is computed with non-aligned phoneme transcriptions.
In this paper 5 low resource languages each with 1-hour training data are used, including Spanish (es), French (fr), Italian (it), Russian (ru) and Tatar (tt).
We report the phone error rate (PER) on the test sets of 5 languages.

\noindent\textbf{Implementation and training details.}
The backbone is VGG-Transformer~\cite{mohamed2019transformers} with 2 VGG convolutional blocks and 12 transformer blocks with attention dimension 512 and feed-forward dimension 2048.
The projector is a MLP with a single hidden layer, with hidden units 2048 and output dimension 256.
Frame level batch normalization is applied at the hidden layer before the Relu activation.
The input acoustic feature is 83-dimensional filter bank including 3-dimensional pitches.
VGG layers have a downsampling factor of 2 so that the output embedding is 20ms per frame.
Due to our computational resource, we use the 45M model, which is half smaller than that in~\cite{conneau2020unsupervised,wang2021unispeech}.

Training involves two schemas, supervised pretraining (T-100) to minimize frame averaged cross entropy and unsupervised training to minimize the similarity loss (Eqn.~\ref{eq:loss}).
Adam optimizer is used for training.
Training schedule is denoted as $(N_e, lr, w_1, w_2, w_3)$, where $N_e$ is the total training epochs, $lr$ is the maximum learning rate with warming up, holding and exponentially decaying periods of $w_1$, $w_2$, $w_3$ respectively.
We set $(100, 10^{-3}, 20\%, 0, 80\%)$ for T-100, $(50, 10^{-3}, 20\%, 0, 80\%)$ for ST-xx series, and $(50, 5\times10^{-4}, 0, 50\%, 50\%)$ for XLST-multi.
XLST-Mono uses the same schedule as XLST-multi but with different training epochs, varying from 50 to 200 depending on the data size of each language.
Experiments is carried out on 12 GPUs (V100,12G).
For instance, to obtain XLST-multi, the total training time is about 64 hours.

\noindent\textbf{Sequence augmenter.}
Augmenter mentioned in Section~\ref{ssec:XLST} is used in both training schemas.
Specifically, in frequency dimension, following~\cite{Park2019SpecAugment}, 2 windows ranging from 0 to 27 are randomly selected to be masked;
In time dimension, we follow the masking strategy in~\cite{Baevski2020wav2vec2} but with mask length 10 (100ms) and total masking proportion 40\% in a sequence;
For mixup~\cite{zhang2017mixup}, two input sequences are mixed and padded to the same length for the \textit{Main Network} while their targets are individually drawn by the \textit{Target Network} (use true labels in supervised pretraining).
Following~\cite{zhang2017mixup}, the mixing weight is sampled by Beta distribution $\beta\sim Beta(\alpha,\alpha)$.
Here $\alpha$ is set to 1.0.
All above three kinds of augmenter are used at supervised pretraining stage.
And during the unsupervised pretraining stage, for convenience only masking in time and frequency dimension is applied on the \textit{Main Network}.

\begin{table*}[htbp]
	\centering
	\caption{General results. Data used directly to pretrain the model, or used at all stages is listed respectively, * indicates annotated data.}
	\resizebox{\textwidth}{!}{%
	\renewcommand\arraystretch{1.2}
	\begin{tabular}{lcccccccccc}
		\hline
		\multirow{2}{*}{Model}      & \multicolumn{1}{c|}{\multirow{2}{*}{Size}} & \multicolumn{1}{c|}{\multirow{2}{*}{Initialization}} & \multicolumn{1}{c|}{\multirow{2}{*}{Pretraining Data}} & \multicolumn{1}{c|}{\multirow{2}{*}{All Data Used}} & \multicolumn{6}{c}{Downstream PER(\%)}                       \\
		& \multicolumn{1}{c|}{}                        & \multicolumn{1}{c|}{}                                & \multicolumn{1}{c|}{}                                  & \multicolumn{1}{c|}{}                               & es   & fr   & it   & ru   & tt                        & avg  \\ \hline
		\multicolumn{11}{l}{\textit{Results reported in}~\cite{wang2021unispeech}}                                                                                                                                                                                 \\ \hline
		CTC-Transfer                & \multicolumn{1}{c|}{90M} & \multicolumn{1}{c|}{-}         & \multicolumn{1}{c|}{En$_{1350}^{*}$}      	& \multicolumn{1}{c|}{En$_{1350}^{*}$}             	& 12.6 & 16.7 & 16.4 & 17.5 & \multicolumn{1}{c|}{11.2} & 14.9 \\
		UniSpeech           		& \multicolumn{1}{c|}{90M} & \multicolumn{1}{c|}{-}         & \multicolumn{1}{c|}{En$_{1350}^{*}$}        	& \multicolumn{1}{c|}{En$_{1350}^{*}$}             	& 10.9 & 14.8 & 15.2 & 16.1 & \multicolumn{1}{c|}{9.6}  & 13.3 \\
		UniSpeech$^+$              	& \multicolumn{1}{c|}{90M} & \multicolumn{1}{c|}{-}         & \multicolumn{1}{c|}{En$_{1350}^{*}$ + Mono } 	& \multicolumn{1}{c|}{En$_{1350}^{*}$ + Mono}      	& 5.7  & 7.9  & 8.1  & 8.6  & \multicolumn{1}{c|}{6.0}  & 7.3  \\ \hline
		\multicolumn{11}{l}{\textit{Results reported in}~\cite{conneau2020unsupervised}}                                                                                                                                                                           \\ \hline
		XLSR-English                & \multicolumn{1}{c|}{90M} & \multicolumn{1}{c|}{-}         & \multicolumn{1}{c|}{En$_{1350}$}        		& \multicolumn{1}{c|}{En$_{1350}$}             		& 13.7 & 20.0 & 19.1 & 18.6 & \multicolumn{1}{c|}{11.5} & 16.6 \\
		XLSR-Monolingual            & \multicolumn{1}{c|}{90M} & \multicolumn{1}{c|}{-}         & \multicolumn{1}{c|}{Mono}            			& \multicolumn{1}{c|}{Mono}                 		& 6.8  & 10.4 & 10.9 & 11.6 & \multicolumn{1}{c|}{21.4} & 12.2 \\
		XLSR-10                     & \multicolumn{1}{c|}{90M} & \multicolumn{1}{c|}{-}         & \multicolumn{1}{c|}{En$_{557}$ +  Multi$_{793}$}   & \multicolumn{1}{c|}{En$_{557}$ +  Multi$_{793}$}          		& 9.4  & 14.2 & 14.1 & 11.0 & \multicolumn{1}{c|}{7.6}  & 11.3 \\ \hline
		\multicolumn{11}{l}{\textit{English pretrained models (Ours)}}                                                                                                                                                                                             \\ \hline
		T-100                       & \multicolumn{1}{c|}{45M} & \multicolumn{1}{c|}{-}         & \multicolumn{1}{c|}{En$_{100}^{*}$}         	& \multicolumn{1}{c|}{En$_{100}^{*}$}              	& 14.6 & 18.9 & 19.2 & 19.4 & \multicolumn{1}{c|}{12.5} & 16.9 \\
		ST-en457                    & \multicolumn{1}{c|}{45M} & \multicolumn{1}{c|}{T-100}     & \multicolumn{1}{c|}{En$_{457}$}         		& \multicolumn{1}{c|}{En$_{100}^{*}$ + En$_{457}$}  & 12.9 & 16.4 & 17.1 & 17.7 & \multicolumn{1}{c|}{10.9} & 15.0 \\
		ST-en1250                   & \multicolumn{1}{c|}{45M} & \multicolumn{1}{c|}{ST-en457}  & \multicolumn{1}{c|}{En$_{1250}$}        		& \multicolumn{1}{c|}{En$_{100}^{*}$ + En$_{1250}$} & 11.6 & 15.5 & 16.1 & 16.9 & \multicolumn{1}{c|}{10.1} & 14.1 \\ \hline
		\multicolumn{11}{l}{\textit{Monolingual pretrained models (Ours)}}                                                                                                                                                                                           \\ \hline
		XLST-Mono (MA)              & \multicolumn{1}{c|}{45M} & \multicolumn{1}{c|}{ST-en1250} & \multicolumn{1}{c|}{Mono}            			& \multicolumn{1}{c|}{En$_{100}^{*}$ + En$_{1250}$ + Mono}			& 6.6  & 8.9  & 9.7  & 9.9  & \multicolumn{1}{c|}{8.2}  & 8.7  \\ \hline
		\multicolumn{11}{l}{\textit{Multilingual pretrained models (Ours)}}																																																			\\ \hline
		XLST-Multi (MA, $unbalanced$) & \multicolumn{1}{c|}{45M} & \multicolumn{1}{c|}{ST-en457}  & \multicolumn{1}{c|}{Multi$_{793}$}      	& \multicolumn{1}{c|}{En$_{100}^{*}$ + En$_{457}$ +  Multi$_{793}$} & 7.4  & 8.9  & 10.7 & 12.4 & \multicolumn{1}{c|}{8.0}  & 9.5  \\
		XLST-Multi (MA)             & \multicolumn{1}{c|}{45M} & \multicolumn{1}{c|}{ST-en457}  & \multicolumn{1}{c|}{Multi$_{793}$}      		& \multicolumn{1}{c|}{En$_{100}^{*}$ + En$_{457}$ +  Multi$_{793}$} & 7.6  & 9.9  & 10.4 & 11.1 & \multicolumn{1}{c|}{6.9}  & 9.2  \\ \hline
		\end{tabular}%
	}
	\label{Tab:general results}
\end{table*}

\section{Results \& analysis}
\subsection{General results}
The general results of our pretrained models as well as recent reported baselines are shown in Table~\ref{Tab:general results}.
The reported systems use similar model architectures as ours, which consist of a CNN front following by a Transformer backbone.
Table~\ref{Tab:general results} shows that:

1) Even English pretrained models show good transferability on other languages.
This may be contributed to the effectiveness of supervised pretraining (T-100), as well as self-training that explores more information from large-scale un-annotated data (ST-en1250).
Specifically, with 100 hours of annotated data, ST-en1250 achieves 15.1\% relatively PER reduction against Wav2vec 2.0 English model (XLSR-English), and even 5.4\% better than fully supervised CTC model (CTC-Transfer).

2) Multilingual pretrained models significantly improve the performance of downstream ASR.
In Table~\ref{Tab:general results}, 34.7\% PER reduction (XLST-multi) over the best English pretrained model (ST-en1250) is observed.
This indicates that even starting from a English model, XLST can still learn important knowledge of other languages, as long as un-annotated data from other languages is provided.
What's more, leveraging additional 100h annotated English data, our XLST model (XLST-multi) achieves 18.6\% relatively PER reduction over Wav2vec 2.0 multilingual model (XLSR-10).

3) Monolingual pretrained models (XLST-mono) outperform the best English model with a significant margin by effectively learning the language-specific knowledge.
Compared with XLSR-monolingual, one could see that English knowledge (as well as data) is important for other languages, especially with extreme low resource.
XLST-mono performs worse than Unispeech$^+$~\cite{wang2021unispeech} perhaps due to the smaller model and much less annotated data used for pretraining.

\begin{table}[htb]
	\centering
	\caption{Effectiveness of moving average (MA) mechanism.}
	\renewcommand\arraystretch{1.2}
	\begin{tabular}{lcc}
		\hline
		Model                   	& \multicolumn{1}{l}{Initialization} & \multicolumn{1}{c}{Avg downstream PER (\%)} \\ \hline
		ST-en1250               	& ST-en457                           & 14.1                        \\ \hline
		ST-en1250               	& T-100		                         & 14.5                        \\
		\quad+ MA ($\lambda=0.9998$) & T-100                              & 14.8                        \\ \hline
		XLST-multi (unbalanced) 	& ST-en457                           & 12.2                        \\
		\quad+ MA ($\lambda=0.9999$) & ST-en457                           & 9.5                         \\ \hline
	\end{tabular}
	\label{Tab:MA}
\end{table}

\subsection{Ablation study}
To further investigate what influence the XLST framework, we do ablation experiments as follows.

\noindent\textbf{Moving average (MA).}
Table~\ref{Tab:MA} shows the average downstream PER of 5 languages from different pretrained models trained with or without the moving average (MA) mechanism.
For MA, the parameter $\lambda$ is searched from $(0.9996, 0.9998, 0.9999)$ according to the average PER on 5 validation sets.
It is observed that in ST-xx setting (model pretrained only with English data), we didn't get better performance with MA.
Simply offline assigning the \textit{Main Network} to the \textit{Target Network} (\textit{i.e.} XLST-en457 to initialize XLST-en1250) performs better.
While in cross-lingual setting, MA brings surprising performance improvement.
We consider this phenomenon due to the language mismatch where the initial English model produces worse targets on other languages, thus online refining them becomes crucial.

\noindent\textbf{Sequence augmenter.}
Since augmenter plays similar role in both supervised and unsupervised pretraining in XLST, we conduct ablation study only on T-100.
The results are shown in Table~\ref{Tab:augmenter}.
All three kinds of augmenter can improve the model's transferability.
Among them, the span masking in time dimension is most effective.
It may be due to that the model is forced to learn stronger context dependency by predicting the information of a masked frame from other frames in a sequence.

\begin{table}[htb]
	\centering
	\caption{Effectiveness of sequence augmenter.}
	\renewcommand\arraystretch{1.2}
	\begin{tabular}{lc}
		\hline
		Model					         			& \multicolumn{1}{l}{Avg downstream PER (\%)} \\ \hline
		T-100                             			& 16.9                        \\
		- without mixup                         	& 18.5                        \\
		- without mixup, F-span-mask            	& 19.8                        \\
		- without mixup, F-span-mask, T-span-mask 	& 25.7                        \\ \hline
	\end{tabular}
	\label{Tab:augmenter}
\end{table}

\section{Conclusion}
In this work, we proposed XLST, a weakly supervised multilingual pretraining framework.
Starting with a prior model trained from non-target high-resource languages, XLST is able to learn multilingual representations with un-annotated data.
Experiments showed that by additionally leveraging just a small amount of annotated data from English, XLST significantly outperformed self-supervised methods in multilingual setting, even with smaller model size.
Moving average mechanism and multi-view augmentation are shown crucial to XLST.
In the future, more experiments with larger model and datasets as well as more elaborate sequence augmenter will be further investigated.

\clearpage
\bibliographystyle{IEEEtran}
\bibliography{refs}
\end{document}